\documentclass{PoS}

\PoS{PoS(BDMH2004)038}
\usepackage{psfig}

\title{The properties of ultra-compact dwarf galaxies and their possible origin}

\ShortTitle{Ultra-compact dwarf galaxies}

\author{\speaker{Michael Hilker}\thanks{Most of our work on UCDs is done in
collaboration with: L.~Infante, M.J.~Drinkwater, M.D.~Gregg, K.~Bekki, 
W.J.~Couch, H.C.~Ferguson, J.B.~Jones, A.M.~Karick, S.~Phillipps, \& 
M.J.~West}\\
        Sternwarte der Universit\"at Bonn, Germany\\
        E-mail: \email{mhilker@astro.uni-bonn.de}}

\author{Steffen Mieske\\
        Sternwarte der Universit\"at Bonn, Germany\\
        E-mail: \email{smieske@astro.uni-bonn.de}}

\abstract{In this contribution the discovery and properties of ultra-compact 
dwarf galaxies are presented, and their possible origin is discussed. This new 
type of galaxy resides in the cores of galaxy clusters. In the fundamental 
plane diagram of stellar systems, the luminosity and kinematical and structural 
properties of ultra-compact dwarfs locate them inbetween globular clusters and 
small compact ellipticals (like M32). This regime would also be occupied by 
nuclei of dwarf ellipticals if they were isolated from their galactic envelope 
or by the merger product of super star-cluster complexes as they are found in
strongly interacting galaxies. Future investigations of ultra-compact dwarfs
in different environments have to show which of these formation scenarios is to
be favoured.}

\FullConference{Baryons in Dark Matter Halos\\
		 5-9 October 2004\\
		 Novigrad, Croatia}

\begin{document}

\begin{centering}
\begin{figure}
\psfig{figure=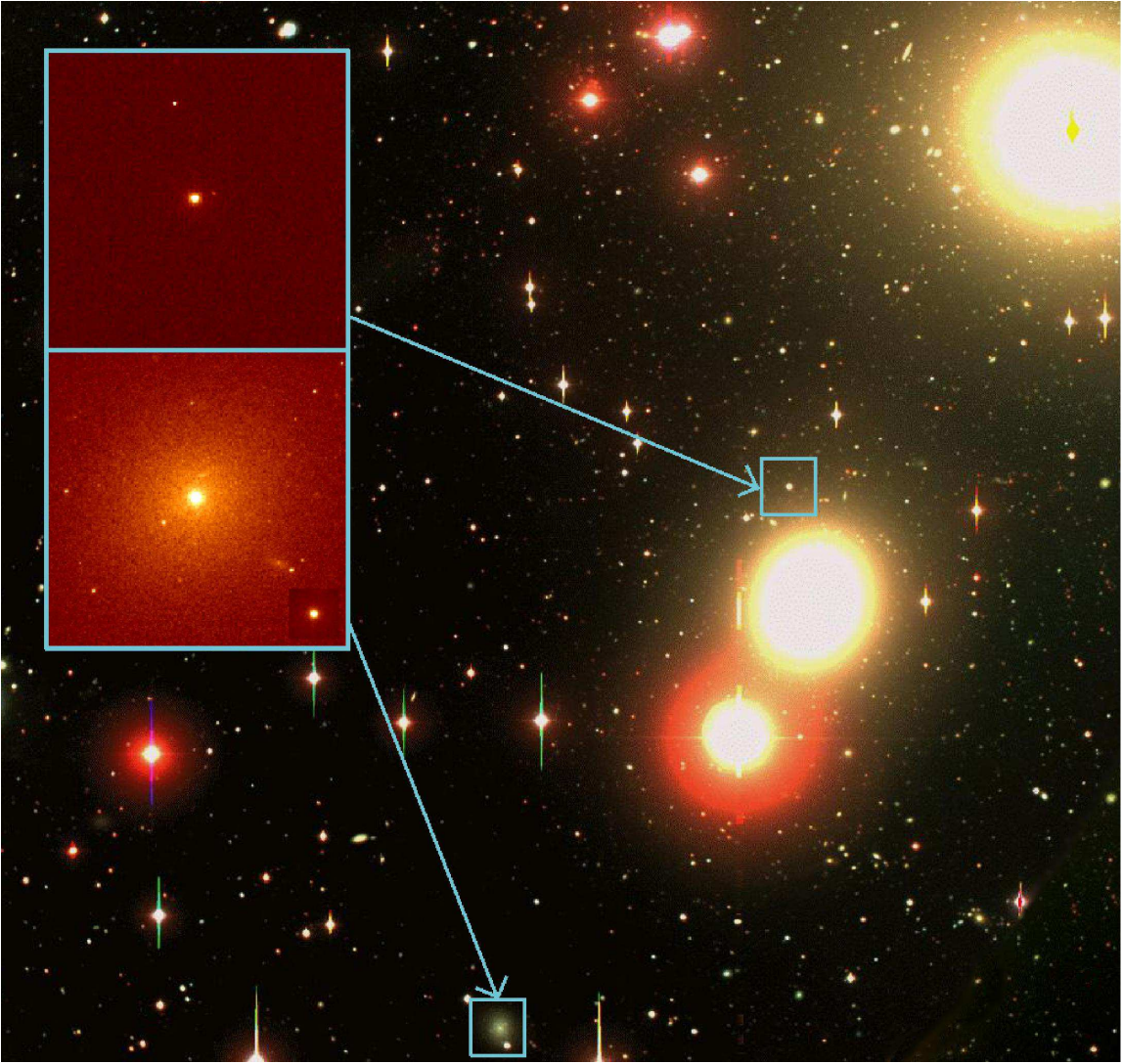,width=15.2cm}
\caption{Central region of the Fornax cluster with the giant ellipticals
NGC~1399 (upper right corner) and NGC~1404. The inlays show an ultra-compact
dwarf galaxy (UCD) and a nucleated dwarf elliptical. This image is a 3-colour
composite ($BVI$) and was taken with the 2.5m DuPont telescope at Las Campanas 
Observatory in order to search for dwarf spheroidals in the cluster (Hilker et 
al. 2003).}
\label{fornax}
\end{figure}
\end{centering}

\section{The discovery of ultra-compact dwarf galaxies}

Ultra-compact dwarf galaxies (UCDs) have recently been proposed as a new
galaxy type (Drinkwater et al.~2003). They have first been discovered in the 
Fornax cluster (see Fig.~\ref{fornax}).

In 1999, Hilker et al. presented the results of a spectroscopic survey 
of selected objects in the cluster core and reported on the 
confirmation of two new ``very compact'' members. They found them to ``have 
photometric properties that can be explained by a very bright globular cluster 
as well as by a compact elliptical like M32''. Also they suggested that these
objects might ``represent the nuclei of dissolved dwarf ellipticals'', and 
stated that ``it would be interesting to investigate, whether there are more 
objects of this kind hidden among the high surface brightness objects in the 
central Fornax cluster''.
Indeed, one year later the results of a systematic spectroscopic survey
in the central two degrees of the Fornax cluster (FCSS: Fornax Cluster
Spectroscopic Survey, Drinkwater et al.~2000a) were published.
Three further compact objects in the luminosity range $-12.2<M_V<-11.8$ mag
were discovered (Drinkwater et al.~2000b).
This survey is unique in the sense that {\it all} objects in the two degree
field, resolved as well as unresolved, have been targeted.
Due to their compactness the five new Fornax members objects were named 
``ultra-compact dwarfs'' (UCDs, Phillipps et al.~2001).

The absence of UCDs in the outer parts of the FCSS field suggests that
these objects constitute a galaxy population that is preferentially found in
the dense central region of galaxy clusters. No counterpart of a Fornax UCD
has been found in the Local Group so far.

\begin{figure}
\psfig{figure=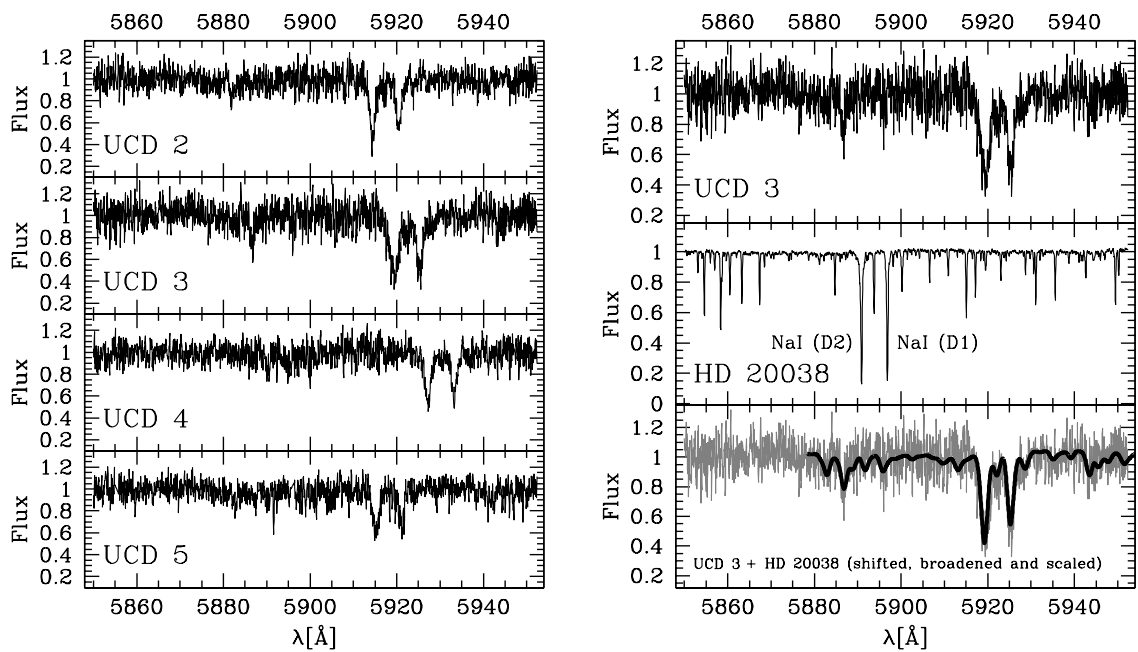,width=14.8cm,bbllx=15mm,bblly=144mm,bburx=195mm,bbury=246mm}
\caption{High resolution echelle spectra (UVES) that have been used for the 
determination of the internal velocity dispersion of the UCDs. On the left,
the spectral region around the Na doublet is shown for four UCDs. On the right,
the spectrum of UCD~3, a standard star, and the shifted, broadened and scaled 
spectrum of the standard star (grey curve) superimposed onto the UCD spectrum 
are shown.}
\label{spectra}
\end{figure}

\section{Properties of ultra-compact dwarfs}

\begin{figure}
\psfig{figure=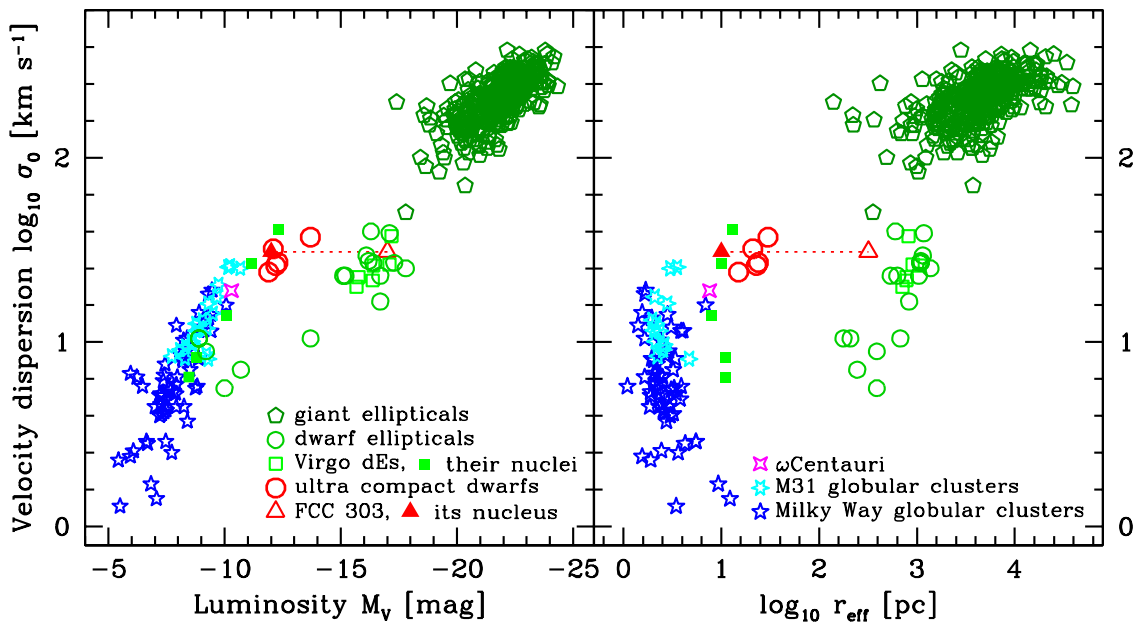,width=14.5cm,bbllx=13mm,bblly=142mm,bburx=195mm,bbury=246mm}
\caption{Fundamental plane, absolute magnitude $M_V$ vs. central velocity
dispersion $\sigma_0$ (left) and effective radius $r_{\rm eff}$ vs. $\sigma_0$
(right), for the indicated stellar systems. A scenario in which the stellar
envelopes of nucleated dwarf galaxies are disrupted can explain
the observed properties of the UCDs. See, for example, the location of FCC~303
and its nucleus (connected by a dashed line).}
\label{fpplot}
\end{figure}

UCDs resemble globular clusters in their general appearance, but are up to 100 
times more massive 
(1-5$\times10^7 M_{\odot}$) and slightly more extended ($r_{\rm eff}\leq 
30$pc). Their luminosities are comparable to those of nuclei of dwarf 
ellipticals ($-13.5<M_V<-11.0$), their photometric colours to those of 
metal-rich bulge globular clusters of giant ellipticals. From high resolution 
spectroscopy of four UCDs in the Fornax cluster, the internal velocity 
dispersion of their stars has been derived (see Fig.~\ref{spectra}). 
Values found range from 20 to 40 km~s$^{-1}$. Using the sizes of the UCDs -- 
derived from high resolution surface brightness profiles 
-- their masses and mass-to-light ratios ($M/L_B$) could be estimated. These 
are in the range of 2-4 in solar units (Drinkwater et al.~2003).
This is slightly higher than the $M/L$ ratio of globular clusters, but much 
lower than found for dwarf spheroidal galaxies of similar mass. 
In the fundamental plane diagram of stellar systems (luminosity vs. velocity 
dispersion), the measured values of the UCDs are lying slightly off the 
relation for globular clusters, but are consistent with an extrapolation to 
fainter luminosities of the Faber-Jackson relation for elliptical galaxies 
(see Fig.~\ref{fpplot}). 
Other compact objects that are found at about the same location 
in the fundamental plane are nuclei of dwarf ellipticals (Geha et al. 2002),
the nuclear clusters of late-type spirals (Walcher et al. 2004), as well as
the young super star cluster W3 in NGC~7252, after evolving over a few Gyr
(Maraston et al. 2003).

\begin{centering}
\begin{figure}
\psfig{figure=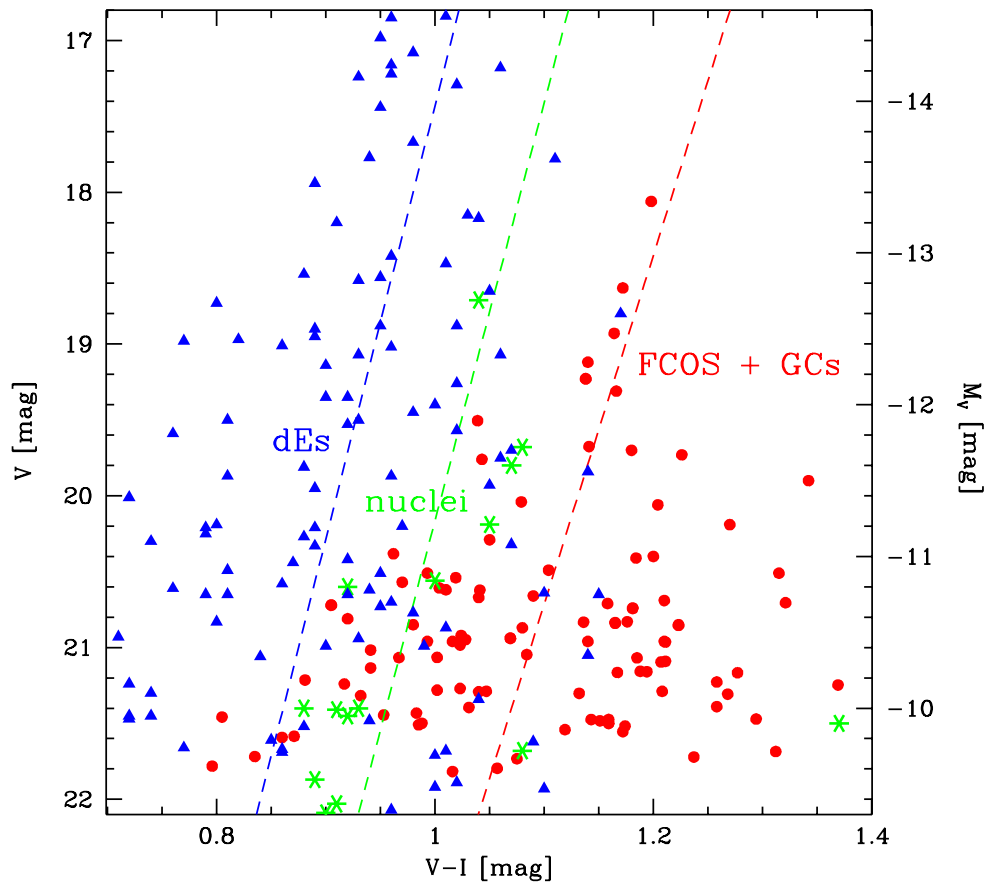,width=12.0cm,bbllx=13mm,bblly=68mm,bburx=195mm,bbury=246mm}
\caption{Colour magnitude diagram of dwarf ellipticals (blue triangles, from
Hilker et al.~2003), nuclei of dEs (green asterisks, from Lotz et al.~2004), 
and compact objects (red dots, including bright globular clusters, from Mieske 
et al.~2004 and Dirsch et al. 2003). The lines are fits to the data points in 
the ranges $15.0<V_{\rm dEs}<22.0$, $V_{\rm nuclei}<24.0$, and 
$V_{\rm compacts}<21.0$.}
\label{cmdall}
\end{figure}
\end{centering}

Bright UCDs ($M_V<-12.0$) do not seem to exist in large numbers in galaxy 
clusters. The fainter ones can easily be confused with the bright globular 
clusters of the extraordinary rich globular cluster systems of the brightest
cluster galaxies (i.e. Dirsch et al.~2003), and therefore their exact 
abundances are unclear.
In order to investigate possible distinctions between UCDs and
bright globular clusters, a systematic spectroscopic survey (Fornax Compact 
Object Survey: FCOS) of the brightest star clusters in the centre of the Fornax
cluster was performed by Mieske et al. (2002, 2004). 54 new 
compact Fornax members were found. Bright compact objects 
($V<20$ or $M_V<-11.4$ mag), including the UCDs, have a higher mean radial
velocity than faint compact objects ($V>20$ mag) at $2\sigma$ significance. 
The mean radial velocity of 
the bright compact objects is consistent with that of the dwarf galaxy 
population in Fornax, but inconsistent with that of the central galaxy 
NGC~1399's globular cluster
systems. The compact objects show a trend of redder colour with increasing 
luminosity with a suggested slope similar to that of the well known 
colour-magnitude relation of dEs, but shifted about 0.2 mag redwards (see
Fig.~\ref{cmdall}). 
The brightness distribution of compact objects observed in the FCOS shows a soft
transition between UCDs and GCs with a slight overpopulation with respect to 
the extrapolated very bright end of NGC 1399's GC luminosity function.
The spatial distribution of bright compact objects within
the cluster is, in comparison to the faint ones, more extended.
Fainter than $V\simeq$ 20 mag, the
majority of the objects seem to be dominated by genuine GCs.

\section{Scenarios for the origin of UCDs}

\begin{centering}
\begin{figure}
\psfig{figure=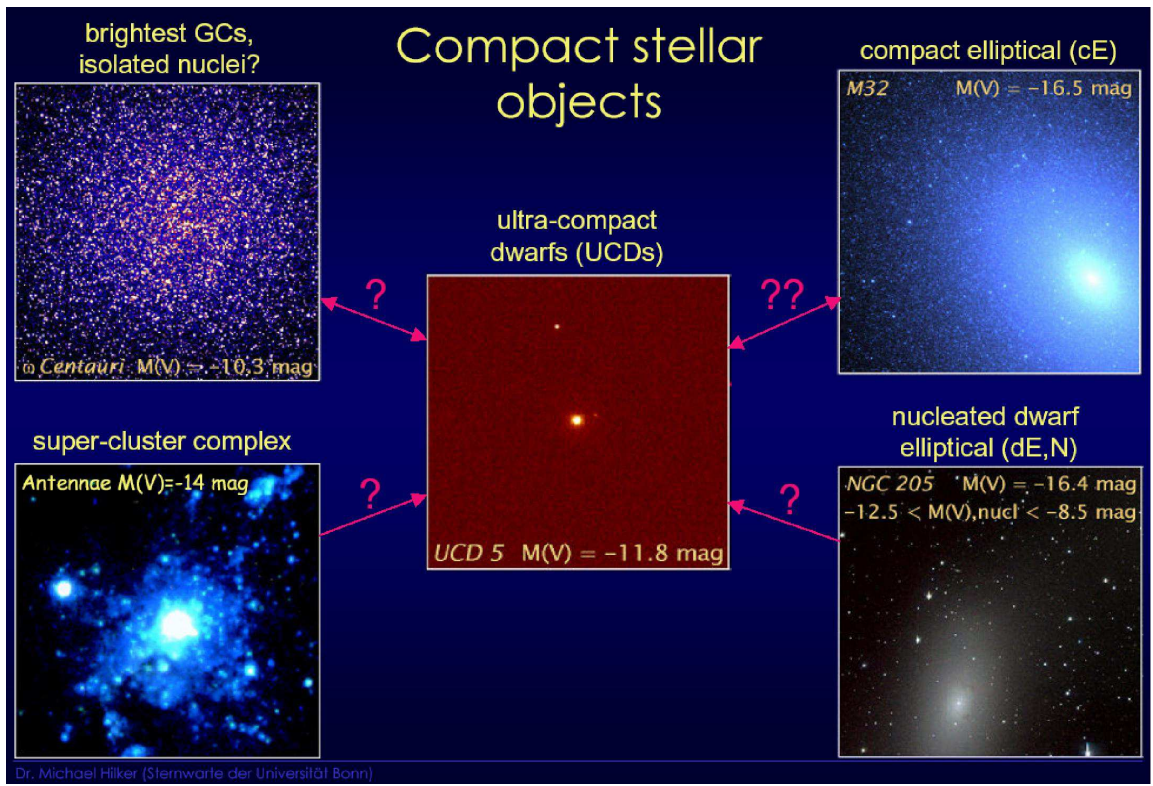,width=15.2cm}
\caption{Compact stellar objects in comparison. How were
the ultra-compact dwarfs formed? Are they genuine globular clusters, just more
massive? Were they
formed from merging super star clusters? Or the low mass end of compact
ellipticals? Mostly favoured is a scenario in which they represent the
isolated nuclei of threshed dwarf galaxies. Note that the objects shown are
located at different distances.}
\label{compacts}
\end{figure}
\end{centering}

Apart from the possibility that ultra-compact dwarfs are very extreme 
``ordinary'' globular clusters, various formation scenarios have been 
brought forward to explain their origin and evolution see overview in 
Fig.~\ref{compacts}). Two of
them seem to be most promising:
first, UCDs might be the remnant nuclei of dwarf galaxies that have been
disrupted in the cluster environment (Bekki et al.~2003). Second, UCDs might
have formed from the agglomeration of many young, massive star clusters that
were created during an ancient merger event (Fellhauer \& Kroupa~2002). 
Numerical simulations have shown that the structural and
kinematical properties of UCDs (i.e. their position in the fundamental plane)
can be reproduced by both scenarios.

The findings obtained from the FCOS data (see previous section) suggest that 
a substantial fraction of compact Fornax members brighter than $V\simeq$ 20 
mag might have been created by threshing dE,Ns (Bekki et al. 2003). 
The redder colour-magnitude relation of UCDs as compared to dEs would be 
expected from the stripping scenario if the stripped nuclei trace the 
colour-magnitude relation of their brighter progenitors. Also the number of
UCDs in the Fornax cluster is roughly consistent with the predictions of Bekki 
et al. (2003) for stripped nuclei.

\section{Future research on UCDs}

In order to get further insight into the nature of UCDs more observations are
needed. Three main questions should be addressed: 1) what are the ages and 
metallicities of their stellar populations? 2) what is the internal velocity 
dispersion of nuclei, UCDs, and bright globular clusters? Do they split up in
the fundamental plane? and 3) what is the frequency of UCDs in different
galactic environments? Do they exist in the field?

Concerning the first question, the spectroscopic measurement of line indices
for dwarf nuclei and UCDs would solve existing age-metallicity degeneracies.
HST imaging of dE,N-nuclei in Fornax/Virgo has shown that they are slightly 
bluer than UCDs (Lotz et al.~2004, and see Fig.~\ref{cmdall}). 
This suggests that UCDs either have higher metallicities and/or older 
integrated stellar populations than the present-day nuclei. The latter 
possibility is consistent with the threshing scenario if the ram-pressure 
stripping associated with the gravitational stripping of the UCD progenitor 
removes a large gas fraction from the nuclear region, thus inhibiting or 
lowering the efficiency of subsequent star formation events in the naked 
nucleus.


To answer the the second question, spatially resolved high resolution spectra 
for nuclei, UCDs, and GCs would be needed. Although challenging, one might also
study the radial trend of their internal velocity dispersion. This should 
clarify whether UCDs are dark matter dominated (and thus of galaxian origin) or
not.

For the third question, large spectroscopic surveys in a variety of galactic
environments are needed. UCD candidates can be pre-selected by photometry.
If tidal forces are the main triggers for the formation of UCDs, they should be
extremely sparse in low density regions and most abundant in the densest and
mostly evolved galaxy clusters.


\begin{thebibliography}{99}
\bibitem{}K.~Bekki, W.J.~Couch, M.J.~Drinkwater, Y.~Shioya, \emph{Galaxy 
 threshing and the origin of ultra-compact dwarf galaxies in the Fornax 
 cluster}, \emph{MNRAS} {\bf 344} (2003) 399 [{\tt astro-ph/0308243}].
\bibitem{}B.~Dirsch, T.~Richtler, D.~Geisler, et al., \emph{The Globular 
 Cluster System of NGC 1399. I. A Wide-Field Photometric Study}, \emph{AJ}
 {\bf 125} (2003) 1908 [{\tt astro-ph/0301223}].
\bibitem{}M.J.~Drinkwater, M.D.~Gregg, M.~Hilker, et al., \emph{A class of 
 compact dwarf galaxies from disruptive processes in galaxy clusters}, 
 \emph{Nature} {\bf 423} (2004) 519 [{\tt astro-ph/0306026}].
\bibitem{}M.J.~Drinkwater, J.B.~Jones, M.D.~Gregg, S.~Phillipps, \emph{Compact 
 stellar systems in the Fornax Cluster: Super-massive star clusters or 
 extremely compact dwarf galaxies?}, \emph{PASA} {\bf 17} (2000b) 227 [{\tt 
 astro-ph/0002003}].
\bibitem{}M.J.~Drinkwater, S.~Phillipps, J.B.~Jones, et al., \emph{The Fornax 
 spectroscopic survey. I. Survey strategy and preliminary results on the 
 redshift distribution of a complete sample of stars and galaxies}, \emph{A\&A}
 {\bf 355} (2000a) 900 [{\tt astro-ph/0001520}].
\bibitem{}M.~Fellhauer, P.~Kroupa, \emph{The formation of ultracompact dwarf 
 galaxies}, \emph{MNRAS} {\bf 330} (2002) 642 [{\tt astro-ph/0110621}].
\bibitem{}M.~Geha, P.~Guhathakurta, R.P.~van~der~Marel, \emph{Internal 
 Dynamics, Structure, and Formation of Dwarf Elliptical Galaxies. I. A 
 Keck/Hubble Space Telescope Study of Six Virgo Cluster Dwarf Galaxies}, 
 \emph{AJ} {\bf 124} (2002) 3073 [{\tt astro-ph/0206153}].
\bibitem{}M.~Hilker, L.~Infante, G.~Vieira, M.~Kissler-Patig, T.~Richtler, 
 \emph{The central region of the Fornax cluster. II. Spectroscopy and radial 
 velocities of member and background galaxies}, \emph{A\&AS} {\bf 134} (1999)
 75 [{\tt astro-ph/9807144}].
\bibitem{}M.~Hilker, S.~Mieske,  L.~Infante, \emph{Faint dwarf spheroidals in 
 the Fornax Cluster. A flat luminosity function}, \emph{A\&A} {\bf 397} (2003)
 L9 [{\tt astro-ph/0212044}].
\bibitem{}J.M.~Lotz, B.W.~Miller, H.C.~Ferguson, \emph{The Colors of Dwarf 
 Elliptical Galaxy Globular Cluster Systems, Nuclei, and Stellar Halos}, 
 \emph{ApJ} {\bf 613} (2004) 262 [{\tt astro-ph/0406002}].
\bibitem{}C.~Maraston, N.~Bastian, R.P.~Saglia, et al., \emph{The dynamical 
 mass of the young cluster W3 in NGC 7252: Heavy-Weight globular cluster or 
 ultra compact dwarf galaxy?}, \emph{A\&A} {\bf 416} (2004) 467 [{\tt 
 astro-ph/0311232}].
\bibitem{}S.~Mieske, M.~Hilker, L.~Infante, \emph{Ultra compact objects in the 
 Fornax cluster of galaxies: Globular clusters or dwarf galaxies?}, \emph{A\&A}
 {\bf 383} (2002) 823 [{\tt astro-ph/0201011}]. 
\bibitem{}S.~Mieske, M.~Hilker, L.~Infante, \emph{Fornax compact object survey 
 FCOS: On the nature of Ultra Compact Dwarf galaxies}, \emph{A\&A} {\bf 418}
 (2004) 445 [{\tt astro-ph/0401610}].
\bibitem{}S.~Phillipps, M.J.~Drinkwater, M.D.~Gregg, J.B.~Jones, 
 \emph{Ultracompact Dwarf Galaxies in the Fornax Cluster}, \emph{ApJ} {\bf 560}
 (2001) 201 [{\tt astro-ph/0106377}].
\bibitem{}C.J.~Walcher, R.P.~van~der~Marel, D.~McLaughlin, et al., \emph{Masses 
 of star clusters in the nuclei of bulge-less spiral galaxies}, \emph{ApJ}
 (2004) [{\tt astro-ph/0409216}].
\end{thebibliography}
\end{document}